\documentclass[aps,prc,twocolumn,showpacs,preprintnumbers,amsmath,amssymb,superscriptaddress,nofootinbib]{revtex4-1}

\usepackage{graphicx}
\usepackage{dcolumn}
\usepackage{bm}
\usepackage[dvipsnames]{xcolor}
\usepackage[pdftex,colorlinks,linkcolor = blue,urlcolor  = blue,citecolor = blue]{hyperref}
\usepackage{epsfig}
\usepackage{dcolumn}
\usepackage{epstopdf}
\usepackage{float}
\usepackage{setspace}
\usepackage{titlesec}

\begin{document}



\title{Independent Normalization for $\gamma$-ray Strength Functions: The Shape Method }

\author{M.~Wiedeking}
\email{wiedeking@tlabs.ac.za}
\affiliation{Department of Subatomic Physics, iThemba LABS, P.O. Box 722, Somerset West 7129, South Africa}
\affiliation{School of Physics, University of the Witwatersrand, Johannesburg 2050, South Africa}

\author{M.~Guttormsen}
\affiliation{Department of Physics, University of Oslo, NO-0316 Oslo, Norway}

\author{A.C.~Larsen}
\affiliation{Department of Physics, University of Oslo, NO-0316 Oslo, Norway}

\author{F.~Zeiser}
\affiliation{Department of Physics, University of Oslo, NO-0316 Oslo, Norway}

\author{A.~G\"{o}rgen}
\affiliation{Department of Physics, University of Oslo, NO-0316 Oslo, Norway}

\author{S.~N.~Liddick}
\affiliation{National Superconducting Cyclotron Laboratory, Michigan State University, East Lansing, Michigan 48824, USA}
\affiliation{Department of Chemistry, Michigan State University, East Lansing, Michigan 48824, USA}

\author{D.~M\"{u}cher}
\affiliation{Department of Physics, University of Guelph, Guelph, Ontario N1G 2W1, Canada}
\affiliation{TRIUMF, 4004 Wesbrook Mall, Vancouver, BC V6T 2A3, Canada}

\author{S.~Siem}
\affiliation{Department of Physics, University of Oslo, NO-0316 Oslo, Norway}

\author{A.~Spyrou}
\affiliation{National Superconducting Cyclotron Laboratory, Michigan State University, East Lansing, Michigan 48824, USA}
\affiliation{Department of Physics and Astronomy, Michigan State University, East Lansing, Michigan 48824, USA}
\affiliation{Joint Institute for Nuclear Astrophysics, Michigan State University, East Lansing, Michigan 48824, USA}

\date{\today}

\begin{abstract}
The Shape method, a novel approach to obtain the functional form of the  $\gamma$-ray strength function ($\gamma$SF) in the absence of neutron resonance spacing data, is introduced. When used in connection with the Oslo method the slope of the Nuclear Level Density (NLD) is obtained simultaneously. 
The foundation of the Shape method lies in the primary $\gamma$-ray transitions which preserve information on the functional form of the $\gamma$SF. The Shape method has been applied to $^{56}$Fe, $^{92}$Zr, $^{164}$Dy, and $^{240}$Pu, which are representative cases for the variety of situations encountered in typical NLD and $\gamma$SF studies. 
The comparisons of results from the Shape method to those from the Oslo method demonstrate that the functional form of the $\gamma$SF is retained regardless of nuclear structure details or $J^\pi$ values of the states fed by the primary transitions. 
\end{abstract}


\maketitle
\section{Introduction}
\label{sec:level1}

The number of nuclear levels per energy interval, the nuclear level density (NLD), and the $\gamma$-ray strength function ($\gamma$SF), which is a measure of the average reduced $\gamma$-ray decay probability, have received significant experimental and theoretical attention over the last decade. 
The necessity for reliable $\gamma$SF data has compelled the International Atomic Energy Agency to establish a dedicated $\gamma$SF database together with recommendations \cite{Goriely2019b}.
The demand for $\gamma$SFs and NLDs is driven in part due to their relevance to astrophysical nucleosynthesis via capture processes \protect\cite{Arnould2007,Mumpower2016,Larsen2019,Arnould2020}. 
Recent experimental results have clearly demonstrated that capture cross sections can be reliably obtained using NLDs and $\gamma$SFs as input into reaction models \cite{Kheswa2015, Spyrou2014, Larsen2016, Malatji2019}, which are based on the Hauser-Feshbach approach \cite{Hauser1952}. 

Several experimental methods exist \cite{Goriely2019b} to extract $\gamma$SFs from experimental data, and of those the Oslo method \cite{Schiller2000} has been extensively used. 
The advantage of the Oslo method lies in its ability to simultaneously extract the $\gamma$SF and NLD from particle-$\gamma$ coincident data. 
The NLD and $\gamma$SF are traditionally normalized by three external parameters: i) the NLD is normalized to the level densities of discrete states at low excitation energies, ii) the NLD at the neutron separation energy ($S_n$) is constrained to the s-wave neutron resonance spacing ($D_0$), and iii) the absolute value of the $\gamma$SF is determined from the average total radiative width of s-wave resonances ($\langle\Gamma_{\gamma {\rm 0}} \rangle$). The functional form of the NLD is linked to that of the $\gamma$SF and can be fully constrained by normalization i) and ii) above. The $\gamma$SFs extracted with the Oslo method have been shown to be reproduced using the alternative $\chi^2$ and Ratio methods, which do not rely on external models or normalization \cite{Wiedeking2012, Krticka2016, Jones2018}. 

Difficulties in normalizing NLD and $\gamma$SF data from the Oslo method emerge for nuclei without available $D_0$ and/or $\langle \Gamma_{\gamma {\rm 0}} \rangle$ values. 
This is the case for many nuclei $A$ when $A-1$ targets are difficult or even impossible to be manufactured, due to the physical or chemical properties of the isotopes and elements, respectively. 
The lack of $D_0$ and $\langle \Gamma_{\gamma {\rm 0}} \rangle$ data present challenges for the normalization of NLDs and $\gamma$SFs. In the absence of normalization data, no coherent prescription is currently available as case-specific approaches \cite{Spyrou2014, Liddick2016, Larsen2016, Kheswa2017, Brits2019} do not appear to be consistently applicable. Even in cases where $D_0$ is known, the normalization procedure introduces a model dependence, which can lead to large uncertainties \cite{Goriely2019b}. 
A reliable approach is highly desirable, especially since the required data needs driven by nucleosynthesis studies primarily involve nuclei for which direct measurements of capture cross sections as well as $D_0$ and $\langle \Gamma_{\gamma {\rm 0}} \rangle$ values are not possible. 
Experimentally, $\gamma$SF and NLD data for nuclei away from the line of stability are readily reachable however, in particular with recent advances in extending the Oslo method to previously inaccessible regions through the $\beta$-Oslo \cite{Spyrou2014, Liddick2016, Liddick2019} and inverse-Oslo \cite{Ingeberg2020} methods. 

In this paper, the {\it Shape method} is introduced, which is a novel and mostly model independent approach to determine the slope of NLDs and $\gamma$SFs extracted with the Oslo method in the absence of measured $D_0$ values. We have also applied the Shape method to $\beta$-decay data on $^{76}$Ge and $^{88}$Kr to explore the extraction of model-independent NLDs away from stability \cite{Mucher2021}.
In section \ref{sec:level2} the Oslo method and the normalization for NLDs and $\gamma$SFs are reviewed. Section \ref{sec:level3} presents the concepts and details of the Shape method, which allows for the normalization of NLDs and $\gamma$SFs. Section \ref{sec:level4} focuses on the Shape method analysis and results on  $^{56}$Fe,  $^{92}$Zr, $^{164}$Dy and $^{240}$Pu. The discussion of results together with recommendations on the use and applicability of the Shape method is provided in section \ref{sec:level5}. Summarizing remarks are made in section \ref{sec:level6}.

\section{Review of the Oslo Method and Normalizations}
\label{sec:level2}

Fermi$^\prime$s golden rule \cite{Fermi} states that the decay rate $\lambda_{if}$ from an initial ($i$) state to a distribution of final ($f$) states is given by a product of the density of final states $\rho_f$ and the transition probability $|\langle f|H'|i \rangle|^{2}$:
\begin{equation}
\lambda_{if} = \frac{2\pi}{\hbar}|\left\langle f| H'|i \right\rangle |^2\rho_f,
\label{101}
\end{equation}

\noindent where $H'$ is the electromagnetic transition operator. 

The Oslo method \cite{Schiller2000} extracts the $\gamma$SF and NLD simultaneously through the following procedure: States in the quasi-continuum (below the particle threshold) are typically populated with charged-particle direct and scattering reactions or following $\beta$ decay. The $\gamma$-ray spectrum is unfolded with the detector response function using an iterative subtraction technique \cite{Guttormsen1996}.  From the unfolded spectra, and with the assumption that the residual nucleus reaches a compound state, the primary $\gamma$-ray spectrum is obtained through the first-generation method \cite{Guttormsen1987}. The first-generation  matrix $P(E_i,E_{\gamma})$ is proportional to the $\gamma$-ray decay probability and can be factorized according to the expression that is derived from Fermi$'$s golden rule (details are found in App.\,C of Ref. \cite{Midtbo2020})

\begin{equation}
P(E_{\gamma},E_i)\propto \rho(E_f)\mathcal{T}(E_{\gamma}),
\label{102}
\end{equation}

\noindent where $\rho(E_f)$ is the nuclear level density and  $\mathcal{T}(E_{\gamma})$ is the transmission coefficient, which is independent of excitation energy ($E_i$) and hence nuclear temperature. This follows from the generalized Brink-Axel hypothesis \cite{Brink1957}, which states that collective excitation modes built on excited states have the same properties as those built on the ground state. The hypothesis 
has been validated in the quasi-continuum with the Oslo method \cite{Guttormsen2016}.
The theoretical matrix $P_{th}(E_{\gamma},E_i)$ is given by \cite{Schiller2000}

\begin{equation}
P_{th}(E_{\gamma},E_i)= \frac{ \rho(E_f)\mathcal{T}(E_{\gamma})}{\sum_{E_\gamma} \rho(E_f)\mathcal{T}(E_{\gamma})}.
\label{104}
\end{equation}

The $\rho(E_f)$ and $\mathcal{T}(E_{\gamma})$ can be simultaneously extracted by performing a $\chi^2$ minimization between the theoretical $P_{th}(E_{\gamma},E_i)$ and experimental $P(E_{\gamma},E_i)$ first-generation matrices \cite{Schiller2000}.

From Eq.~(\ref{104}) an infinite number of solutions are obtained, and the physical solution is found by normalizing $\mathcal{T}(E_{\gamma})$ and $\rho(E_{f})$ to experimental data \cite{Schiller2000} with

\begin{equation}
\tilde{\rho}{(E_f)=A\rho{(E_f)}}e^{\alpha E_{f}},
\label{105}
\end{equation}

\noindent and

\begin{equation}
\tilde{\mathcal{T}}{(E_\gamma)=B\mathcal{T}{(E_{\gamma})}}e^{\alpha E_{\gamma}},
\label{106}
\end{equation}

\noindent where $A$ and $B$ are constants and $\alpha$ is the common slope\footnote{This is an additional slope transforming $\rho{(E_f)}$ and $\mathcal{T}{(E_\gamma)}$ in the same way as for  $\tilde{\rho}{(E_f)}$ and $\tilde{\mathcal{T}}{(E_\gamma)}$. Note however, that the slopes of $\rho{(E_f)}$ and $\mathcal{T}{(E_\gamma)}$, and $\tilde{\rho}{(E_f)}$ and $\tilde{\mathcal{T}}{(E_\gamma)}$ are in general different.}. The slope $\alpha$ and constant $A$ are determined by the NLD of the known discrete states at lower excitation energies and the total NLD at $S_n$. The functional form of $\rho{(E_f)}$ and $\mathcal{T}{(E_\gamma)}$ is defined from the $\chi^2$ fit to the primary $\gamma$-ray matrix $P(E_{\gamma},E_i)$. For a detailed discussion and implementation of the Oslo method, see Ref.\,\cite{Midtbo2020}.

In this work, data from $^{56}$Fe \cite{Larsen2013a},  $^{92}$Zr \cite{Guttormsen2017}, $^{164}$Dy \cite{Renstrom2018}, and $^{240}$Pu \cite{zeiser2019} have been reanalysed with the Oslo method using an intrinsic spin-distribution for the absolute normalization at $S_n$. The $\gamma$SFs of those nuclei may therefore deviate slightly from results presented in previous publications. The form of the spin-distribution is assumed to follow ~\cite{Ericson1960}
\begin{equation}
g(E,J) \simeq \frac{2J+1}{2\sigma^2(E)}\exp\left[-(J+1/2)^2/2\sigma^2(E)\right],
\label{eq:spindist}
\end{equation}
where $E$ is the excitation energy, $J$ the spin, and the spin cutoff parameter $\sigma(E)$ is assumed to have the functional form
\begin{equation}
\sigma^2(E)=\sigma_d^2 + \frac{E-E_d}{S_n-E_d}\left[\sigma^2(S_n)-\sigma_d^2\right],
\label{eq:sigE}
\end{equation}
determined by two excitation energies. At the lower excitation energy $E=E_d$, we determine the spin cutoff parameter $\sigma_d$ from known discrete levels.
The second point at $E=S_n$ is estimated assuming a rigid moment of inertia~\cite{egidy2005,Egidy2006}
\begin{equation}
\sigma^2(S_n) =  0.0146 A^{5/3}  \frac{1+\sqrt{1+4aU_n}}{2a},
\label{eqn:eb}
\end{equation}
where $A$ is the mass number, $a$ is the NLD parameter,
$U_n=S_n-E_1$ is the intrinsic excitation energy,
and $E_1$ is the energy-shift parameter. 

At $S_{n}$, normalization is achieved from NLDs calculated with \cite{Schiller2000}

\begin{equation}
\rho(S_{n}) = \frac{2\sigma^{2}(S_{n})}{D_{0}}  \frac{1}{(J+1)e^{\left[-\frac{(J+1)^{2}}{2\sigma^{2}(S_{n})}\right]} + J e^{\left[ -\frac{J^{2}}{2\sigma^{2}(S_{n})}\right]}}.
\end{equation}

\noindent{}The experimental $D_{0}$ value is obtained from $\ell = 0$ (s-wave) neutron resonance spacing data which are typically retrieved from Refs.~\cite{Capote2009a, Mughabghab2006} and $J$ is the initial spin of the target nucleus. Generally, NLDs can only be extracted to excitation energies well below $S_n$ with the Oslo method. The absolute normalization at $S_n$, which sensitively depends on the spin distribution, is achieved by extrapolating the NLDs using a variety of level density models, such as the back-shifted Fermi-gas \cite{Gilbert1965}, the constant temperature \cite{Ericson1959}, or the Hartree-Fock-Bogoliubov-plus-combinatorial \cite{Goriely2008} models. 

The absolute normalization parameter $B$ in Eq. \ref{106} is obtained by constraining the experimental data to $\langle {\Gamma_{\gamma {\rm 0}}} \rangle$ for s-wave resonances by \cite{Kopecky1990, Midtbo2020}

\begin{equation}
\begin{split}
&\langle\varGamma_{\gamma 0}(S_{n})\rangle=\frac{1}{2\pi\rho(S_{n},J_{t\pm{1/2}},\pi_{t})}\\
&\times  \sum_{J_f}\int_{0}^{S_n} 
B\mathcal{T}(E_\gamma) \rho(S_{n}-E_{\gamma}, J_{f})dE_{\gamma},\\
\end{split}
\label{q9}
\end{equation} 

\noindent{}where $\pi_t$ is the parity of the target nucleus in the (n, $\gamma$) reaction, $J_f$ and $J_t$ are the spins of the levels in the final and target nucleus, respectively. 

The essential parameters used here for the extraction of the NLDs and $\gamma$SFs are listed in Table~\ref{tab:parameters}. More details on the extraction of NLDs and $\gamma$SFs for $^{56}$Fe,  $^{92}$Zr, $^{164}$Dy, and $^{240}$Pu are discussed in Refs. \cite{Larsen2013a, Guttormsen2017,Renstrom2018, zeiser2019}.

\begin{table*}[]
\caption{Parameters used for the extraction of NLDs and $\gamma$SFs (see text for details). }
\begin{tabular}{ccc|cc|cc|c|c|c |c}
\hline
\hline
Nucleus &$S_n$	&  $D_0$    & a$^c$         & $E_1$$^c$  &$E_d$&$\sigma_d$&$\sigma(S_n)$&$\rho(S_n)$ &$T_{\rm CT}$ &   $\langle\Gamma_{\gamma}\rangle$\\
        &[MeV]  & [eV]     	&[MeV$^{-1}$]& [MeV]&[MeV]&          &             &[MeV$^{-1}$]&       [MeV] &            [meV]      \\ \hline
$^{56}$Fe&11.197&   -       &  6.196     & 0.94       & 2.70 &2.5&   4.05    & 2870(680)$^\dagger$&1.35(5)&    1900(600)$^\dagger$        \\
$^{92}$Zr&8.635 &514(15)$^a$	&10.4       & 0.66    	  & 3.0      & 3.0      &   4.50     &   16640(490)&     0.90(2)  & 131(56) \\
$^{164}$Dy&7.658&6.8(6)$^b$    &18.12        &0.31        &1.09      & 3.6  	   &    6.91   & 2.59(52)$\times10^{6}$ & 0.59(2) & 113(13)   \\
$^{240}$Pu&6.534&2.20(9)$^b$    &25.16      & 0.12       & 0.87     & 3.2      &   8.43     & 32.7(66)$\times10^{6}$& 0.44(3)	&  43(4)\\
\hline
\hline
\end{tabular}
\newline
$^\dagger$Estimated from systematics corresponding to \textit{norm-2} in Ref.~\cite{Larsen2017}.
\newline
$^a$ Value from \cite{Mughabghab2006}.
\quad
$^b$ Value from \cite{Capote2009a}.
\quad
$^c$ Values from \cite{egidy2005,Egidy2006}.
\newline
\label{tab:parameters}

\end{table*}

The relationship between $\mathcal{T}(E_{\gamma})$ and the $\gamma$SF ($f_{XL}(E_{\gamma})$) with $XL$ being the type and multipolarity of the radiation, respectively, is \cite{Capote2009a}

\begin{equation}
\mathcal{T}_{XL}(E_{\gamma}) =2\pi {E_{\gamma}}^{2L+1}f_{XL}(E_{\gamma}).
\label{109}
\end{equation} 

\noindent With the assumption that statistical $\gamma$-ray decay is dominated by dipole transitions, the total $\gamma$SF ($f(E_{\gamma})$) becomes

 \begin{equation}
f(E_{\gamma}) =f_{E1}(E_{\gamma})+f_{M1}(E_{\gamma}) =\frac{\mathcal{T}(E_{\gamma})}{2\pi E_{\gamma}^3}.
\label{eq:110}
\end{equation}

The values of
$D_0$ and $\langle {\Gamma_{\gamma {\rm 0}}} \rangle$  from s-wave resonance and to a limited extent $D_1$ and $\langle {\Gamma_{\gamma {\rm 1}}} \rangle$ values from p-wave resonance measurements\footnote{A similar treatment as for $D_0$ can be applied to p-wave neutron resonance spacing data ($D_1$) and if available may be used to provide additional constraints.} are generally available for nuclei which are populated through (n,$\gamma$) reactions on stable targets. For the majority of nuclei the information required by the Oslo method to determine $A$, $B$, and $\alpha$ has not been measured mostly due to the unavailability of targets. This led to many  non-standardized approaches to estimate the values $D_0$ and $\langle  {\Gamma_{\gamma {\rm 0}}} \rangle$ \cite{Spyrou2014, Liddick2016, Larsen2016, Kheswa2017, Brits2019}.

The development of a method with no or only very limited model dependencies, which can be systematically applied to nuclei, is of utmost importance to obtain the normalization when $D_0$ and $\langle  {\Gamma_{\gamma {\rm 0}}} \rangle$ values are not available. 
A new method, the Shape method, will now be described, which provides a prescription for the normalization of the slope of the NLD and $\gamma$SF in the absence of $D_0$. Software for the Oslo and Shape (\texttt{diablo.c}) methods are available from Refs.~\cite{Oslomethod2020, Midtbo2020}.

\section{The Shape Method}
\label{sec:level3}

In this section, the Shape method, which is a technique to obtain the slope of  $\gamma$SF in the absence of measured values of resonance spacing, is presented. The method utilizes concepts from $\gamma$SF measurements using the average resonance proton capture approach and from the Ratio and $\chi^2$ methods using particle-$\gamma$-$\gamma$ coincident data. These approaches are briefly summarized before we continue with a detailed description of the Shape method.

\subsection{Average Resonance Proton Capture}
Experimental data from ($p$,$\gamma$) reactions have been used to deduce the $\gamma$SFs for several $45<A<91$ nuclei for which the proton separation energy ($S_p$) is located below $S_n$ \cite{Goriely2019b}. The methodology is similar to the neutron average resonance capture approach \cite{arc} where several resonances are populated and combined in specific excitation-energy ranges. The use of high-resolution detectors allows for the identification of individual primary $\gamma$-ray transitions to low-lying levels, see for example Refs. \cite{Szeflinski1979, Erlandsson1979}. 
The relative intensities of primary transitions (corrected by $E_{\gamma}^3$), which originate from a given excitation energy region and decay to low-lying levels with the same spin and parity, preserve the shape and hence the energy dependence of the $\gamma$SF.
The proton beam energies, together with the target thicknesses, provide an unambiguous assignment of specific excitation energies. Data of primary transitions to low-lying states of different spins and parities ($J^\pi$) are normalized by weighting the different contributions through the Hauser-Feshbach formalism. Regardless this normalization, the energy dependence of the $\gamma$SF remains completely independent of any model input.

\subsection{Ratio and $\chi^2$ Methods}
The Ratio method \cite{Wiedeking2012} is a model-independent approach to obtain the energy dependence of the $\gamma$SF from correlated particle-$\gamma$-$\gamma$ events following direct reactions. 
The $\gamma$-$\gamma$ coincidence is between the primary $\gamma$-ray transition, originating from the region of the quasi-continuum populated in the reaction, and the transition from low-lying discrete states, which are fed by the primary $\gamma$ rays. When a discrete transition from a low-lying state is detected in coincidence with a charged particle, additional stringent requirements are applied to the primary $\gamma$ ray, so that the energy sum of the discrete and primary transitions is equal to the excitation energy within the energy resolutions of the detectors. 
Any particle-$\gamma$-$\gamma$ event satisfying these conditions provides an unambiguous determination of the origin and destination of the observed primary transition. As long as the primary $\gamma$ rays feed discrete states of the same $J^\pi$ the shape of the $\gamma$SF remains independent of model input by analogy with the (p,$\gamma$) average resonance proton capture method. The ratio $R$ of intensities $N$ for two different primary $\gamma$-ray energies from the same initial excitation energy $E_i$ to discrete low-lying levels of same $J^\pi$ at energies $E_{l_1}$ and $E_{l_2}$ is

\begin{equation}
R = \frac{f(E_i-E_{l_1})}{f(E_i-E_{l_2})} = \frac{N_{l_1}(E_i) (E_i-E_{l_2})^3} 
{N_{l_2}(E_i)(E_i-E_{l_1})^3}.
\label{eq:ratio_method}
\end{equation}

When  the  ratios  from  different  excitation  energies are  compared,  information on the energy dependence of the $\gamma$SF is obtained as demonstrated from (d,p$\gamma\gamma$) \cite{Wiedeking2012}, (p,p'$\gamma\gamma$) \cite{Jones2018}, ($\vec{\gamma},\gamma\gamma$) \cite{Isaak2019}, and (p,$\gamma$) \cite{Scholz2020} reactions.

Data of primary $\gamma$-ray intensities from an excitation energy range to different discrete levels of the same $J^\pi$ and corrected for $E_\gamma^3$, can also be fitted with a $\chi^2$ minimization procedure \cite{Wiedeking2012, Krticka2016, Jones2018}. The set of data from different initial excitation energies are independent of each other and following the $\chi^2$ minimization, which combines the sets from different excitation energy bins, yields information on the shape of the $\gamma$SF. 

\subsection{Shape Method}

In the previous descriptions discrete $\gamma$-ray lines were studied with high-resolution germanium detectors. When the total $\gamma$SF extending across larger excitation and $\gamma$-ray energy ranges is to be measured, the Oslo method with high-efficiency detectors is regularly used. In the following, we will extend the previous techniques and replace the identification of $\gamma$-ray lines from discrete levels $l_j$ with diagonals $D_j$ in a particle-$\gamma$ matrix. 

\begin{figure}[t]
    \includegraphics[clip,width=0.8\columnwidth]{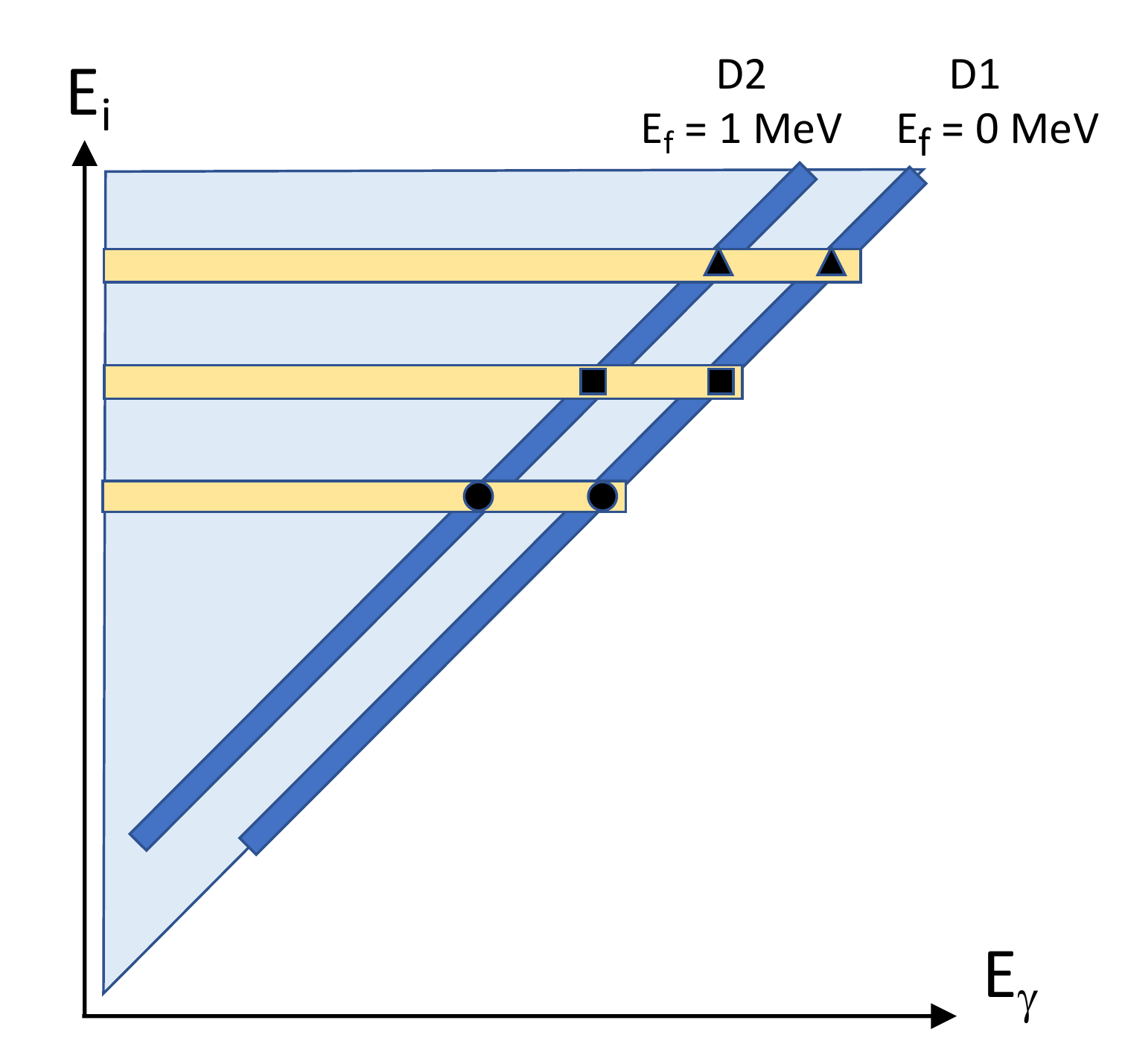}
    \caption{(Color online) Illustration of diagonals (blue) $D_1$ and $D_2$ selecting specific final states in the $P(E_{\gamma}, E_i)$ matrix. Horizontal bars (yellow) indicate three initial excitation energies $E_i$. The number of counts at the crossing points between a diagonal and a bar $(E_{\gamma}, E_i)$ gives the intensity of the $\gamma$ transitions from $E_i$ to $E_i - E_{\gamma}$, here symbolized with filled circles, squares, and triangles. With intensities from two diagonals at the same $E_i$, a pair of internally normalized $\gamma$SF data points can be established.}
    \label{fig:diagonals}
\end{figure}

The diagonals $D_j$ are directly related to the first-generation (or primary) $P(E_{\gamma}, E_i)$ matrix provided by the Oslo method.\footnote{The total $\gamma$-ray matrix (all $\gamma$ rays in a cascade) may be utilized, as long as it is certain that the diagonals contain only primary transitions.} Figure \ref{fig:diagonals} illustrates the concepts of diagonals and symbols used where one may define a final excitation energy $E_f$ fed from an initial excitation energy $E_i$ by a $\gamma$ transition with energy $E_{\gamma}$. This is given by $E_i(E_{\gamma})=E_{\gamma}+E_f$ with $E_f$ fixed and  the diagonals $D_j$ with different $E_f$ are parallel to each other as schematically shown in Fig.\ref{fig:diagonals}. Here, the direct $\gamma$-ray decay from $E_i$ to the ground state is simply given by $E_i(E_{\gamma})=E_{\gamma}$ (within the resolutions of the detectors). 
The diagonals may appear in three variants containing ({\em i}) one final state with given $J^\pi$, ({\em ii}) two or more  specific final states, or in case of high level density, ({\em iii}) a large number of final states (typically $> 20$) with a corresponding average $E_f$ and $J^\pi$. The number of counts along a $D_j$ relates to the $\gamma$SF for a given $E_{\gamma}$ originating from $E_i$. The intensities (counts) given by the content of the pixel $(E_{\gamma}, E_i)$ for two diagonals are exploited to obtain a pair of data points which are proportional to the $\gamma$SF. 

In the following, we assume a symmetric parity distribution with the spin distribution $g(E_i,J_i)$ of Eq.~(\ref{eq:spindist}). Furthermore, we assume the population of a typical state at excitation $E_i$ and spin $J_i$ is given by the cross section $\sigma(E_i,J_i)$.
The number of counts in a diagonal $D_j$ at $(E_{\gamma},E_i)$ with one or more final $J^\pi$ states included, can then be expressed as a sum of products
\begin{equation}
 N_{D}\propto  \sum_{[J_f]}\sum_{J_i=J_f-1} ^{J_i=J_f+1}  \sigma(E_i,J_i)g(E_i,J_i)G(E_i,E_{\gamma},J_i,J_f),
 \label{eq:ndfull}
\end{equation}
where we define $[J_f]$ as the spins of the final levels within the diagonal, e.g.~$[J_f]= [1^-, 2^+, 2^+, 3^-]$ includes the summing of four terms. 
The second sum is restricted to the available $J^\pi$ populated by dipole transitions connecting initial and final states, which generally includes three initial spins. However, in the case of $J_f=0$, only the $J_i=1$ spin is included and for $J_f=1/2$, only the $J_i=1/2$ and $J_i=3/2$ spins are included. 

The third factor $G$ in Eq.~(\ref{eq:ndfull}) is proportional to the $\gamma$-decay width given by
\begin{eqnarray}
\nonumber
&G&(E_i,E_{\gamma},J_i,J_f)\\ 
&\propto& \int_{E_{\gamma} -\Delta /2}^{E_{\gamma} +\Delta/2}{\cal T}(E_i,E_{\gamma}^{\prime},J_i,J_f)\delta(E_i-E^\prime_{\gamma},J_f) dE_{\gamma}^{\prime},
\end{eqnarray}
where $\Delta$ is the energy width of the diagonal which includes the specific final level $J_f$ at $E_f=E_i-E_{\gamma}$. The $\delta$ function assures that one specific level is counted giving $\int \delta\; dE_{\gamma}^{\prime}=1$.
With the assumption that the transmission coefficient is almost constant within this energy bin, it can be placed outside the integral  with a value of ${\cal T}(E_i,E_{\gamma},J_i,J_f)$. 

According to the generalized Brink-Axel hypothesis, the transmission coefficient ${\cal T}(E_i,E_{\gamma},J_i,J_f)$ is assumed to be independent of spin and excitation energy. Thus, we replace the expression for the transmission coefficient by ${\cal T}(E_{\gamma})$, i.e. a function only dependent of $E_{\gamma}$. Furthermore, if we assume the dominance of dipole transitions in the quasi-continuum region, the transmission coefficient can be replaced by the $\gamma$SF through ${\cal T}(E_{\gamma}) = 2\pi f(E_{\gamma})E_{\gamma}^3$ from Eq.~(\ref{eq:110}).

With the considerations above, Eq.~(\ref{eq:ndfull}) can be written as
\begin{equation}
 N_D\propto f(E_{\gamma})E_{\gamma}^3 \sum_{[J_f]}\sum_{J_i=J_f-1} ^{J_i=J_f+1}\sigma(E_i,J_i)g(E_i,J_i).
 \label{eq:ndsimple}
\end{equation}
In the following we will assume that the probability to populate  a certain initial state with spin $J_i$ at a given $E_i$ is approximately independent of spin, i.e.~$\sigma(E_i,J_i) \approx \sigma(E_i,J_i^{\prime})$. 

The Shape method applies for the same $E_i$ but for two different diagonals $D_1$ and $D_2$, see Fig.~\ref{fig:diagonals}. We choose diagonal $D_1$ to represent a lower final excitation energy $E_{f1}$ and $D_2$ a higher final excitation energy $E_{f2}$. At the initial excitation energy $E_i$, the $\gamma$-ray energies are $E_{\gamma_1}= E_i - E_{f1}$ and $E_{\gamma_2}= E_i - E_{f2}$ for diagonals $D_1$ and $D_2$, respectively. 

The strength functions at $E_{\gamma_1}$ and $E_{\gamma_2}$ are determined by the number of counts at the diagonals $D_1$ and $D_2$ for the same initial excitation energy $E_i$, using Eq.~(\ref{eq:ndsimple})
\begin{eqnarray}
 \nonumber
    f(E_{\gamma1})&\propto& \frac{ N_{D1}}{E_{\gamma1}^3 \sum_{[J_{f1}]}\sum_{J_i=J_{f1}-1} ^{J_i=J_{f1}+1}g(E_i,J_i)}\\
    f(E_{\gamma2})&\propto& \frac{ N_{D2}}{E_{\gamma2}^3 \sum_{[J_{f2}]}\sum_{J_i=J_{f2}-1} ^{J_i=J_{f2}+1}g(E_i,J_i)}.
\end{eqnarray}

In synergy with the methods introduced above, such a pair of $\gamma$SF data points is internally normalized and we can determine a $\gamma$SF data-point pair for each $E_i$. The double sum can be omitted if the two diagonals include one final level each of the same $J^\pi$. 
However, such diagonals are often difficult to identify in the data, and it is more common to observe different spins for two diagonals, such as the $0^+$ ground state and the first-exited $2^+$ state in even-even nuclei.

\begin{figure}[t]
    \includegraphics[clip,width=0.8\columnwidth]{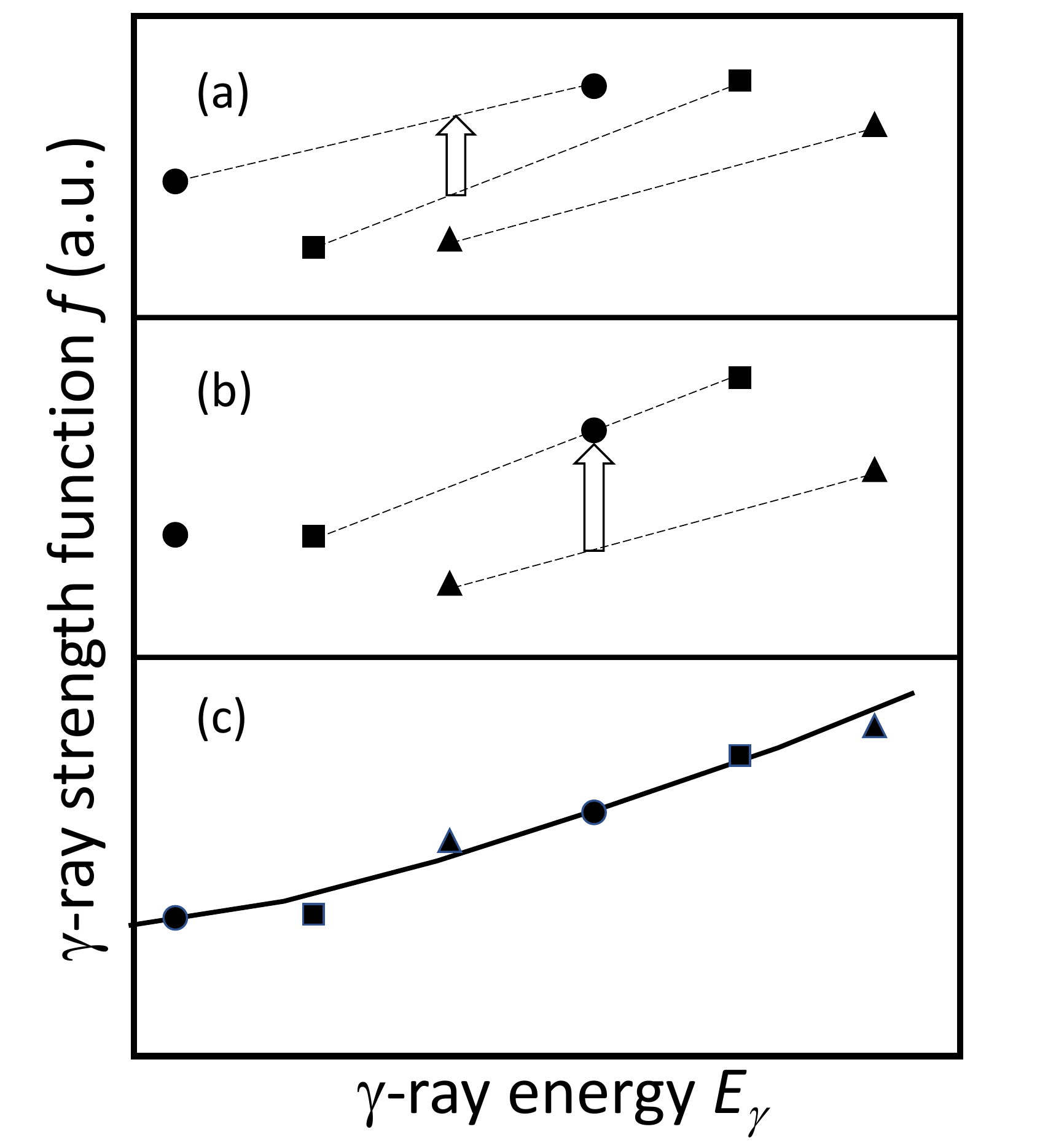}
    \caption{Illustration of the sewing technique for three $\gamma$SF pairs (filled circles, squares and triangles) with each pair connected by dashed lines in (a). The second pair of data points (filled squares) is scaled by a factor to match the first pair of data points at a location indicated by the arrow (filled circles) (a). Then the third pair of data points (filled triangles) is scaled to match the previously corrected data pair (filled squares) at the location of the arrow (b). Finally, the resulting sewed $\gamma$SF is presented in (c) (solid black line).}
    \label{fig:sewing}
\end{figure}

Figure~\ref{fig:sewing} illustrates a sewing technique that allows to connect  pairs of $\gamma$SF data points and is the final step of the Shape method to obtain the functional form of the $\gamma$SF. In this example, we show three different pairs, each from a different $E_i$, marked by filled circle, square and triangle data points. The second and third $\gamma$SF pairs are scaled as explained in the figure caption.
In detail, this is accomplished by finding the average $\gamma$-ray energy $E_{\gamma \rm ave}$ (location of arrow) in between the lowest and highest  $\gamma$SF data points of the two pairs under study. Then we use a logarithmic interpolation of the $\gamma$SF data points for each pair to $E_{\gamma \rm ave}$. 
The resulting sewed $\gamma$SF is represented by the black line to guide the eye in panel (c) and exhibits the shape of the $\gamma$SF. 

\section{Shape Method Analysis and Results}
\label{sec:level4}

In the following, when referring to discrete final levels within the diagonals, we always refer to levels in the data base from the National Nuclear Data Center (NNDC)~\cite{NNDC}. For each application of the Shape Method we use  a  first-generation  matrix  with
$\approx 30-40$ keV/ch on both axes from which the number of counts are determined  through  integration. These  are  then  further compressed into bins of $\approx$ 120 keV/ch unless otherwise noted. Detailed discussions on the comparisons of the results from the Shape and Oslo methods are deferred to Sec. \ref{sec:level5}.

\subsection{Diagonals with the same final $J^\pi$: $^{56}$Fe}

\begin{figure*}[t]
    \includegraphics[clip,width=1.9\columnwidth]{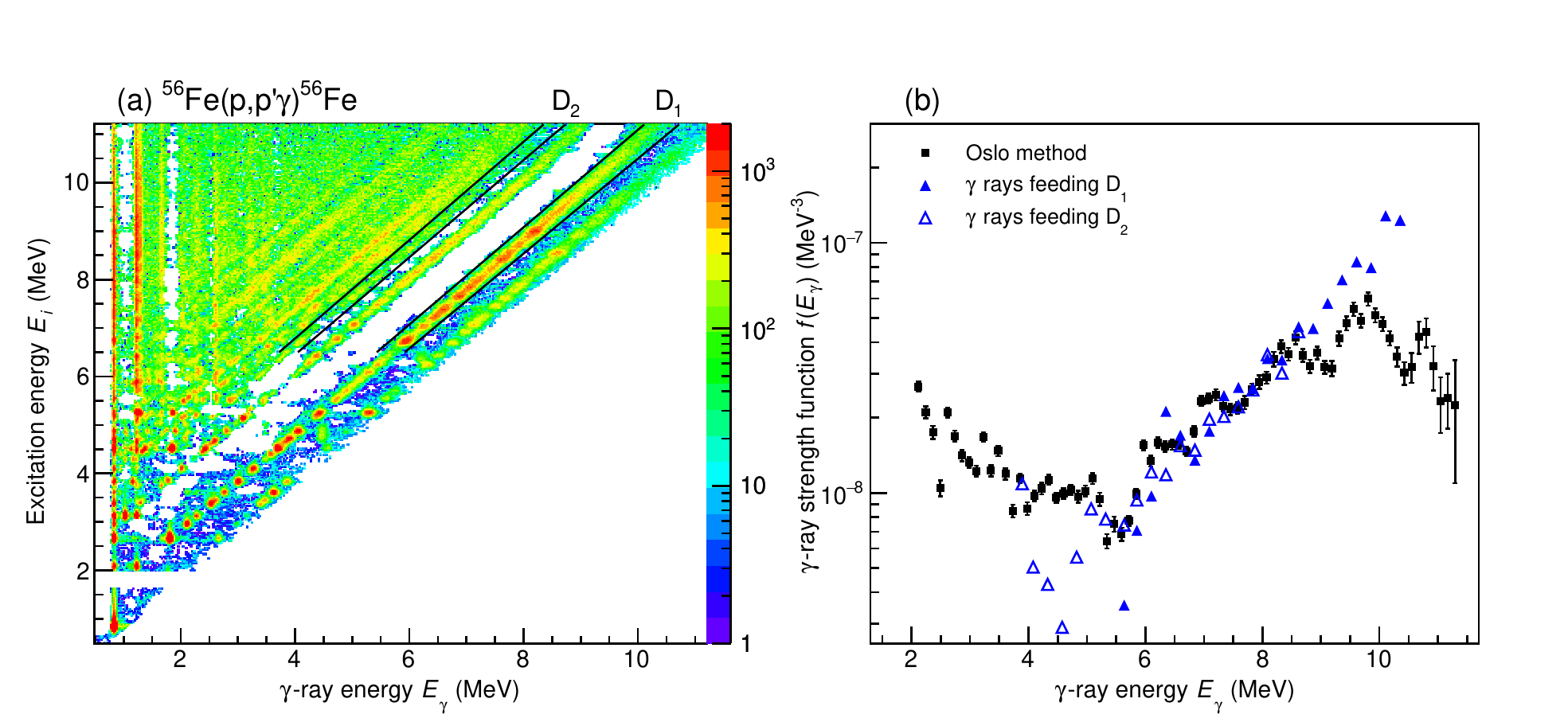}
    \caption{(Color online) (a) The first-generation matrix $P(E_{\gamma}, E_i)$  of $^{56}$Fe showing the cuts on the diagonals decaying to the $2^+_1$ level ($D_1$) at $E_f=847$ keV  and the $2^+_2$ level ($D_2$) at $E_f=2658$ keV. (b) The resulting $\gamma$SF from the Shape method (filled and open blue triangles) compared to the Oslo-method results (solid black squares) ~\cite{Larsen2013a,Larsen2017}. Note that the bin width is 248 keV/ch in this case due to $^{56}$Fe being a relatively light nucleus with a low level density.}
    \label{fig:fg_gsf_56Fe}
\end{figure*}

We utilize data from the $^{56}$Fe($p,p^\prime\gamma$)$^{56}$Fe reaction previously presented in Refs.~\cite{Larsen2013a,Larsen2017}, where the $\gamma$ rays were measured with six large-volume LaBr$_3$(Ce) detectors from the HECTOR$^+$ array~\cite{Giaz2013} and the charged particles with the SiRi silicon telescope \cite{Guttormsen2011}. 
Figure~\ref{fig:fg_gsf_56Fe}a shows the resulting $P(E_\gamma,E_i)$ matrix of $^{56}$Fe. 
Gates were set on the diagonals and correspond to the direct decays to the $2_1^+$ (diagonal $D_1$) and $2_2^+$ (diagonal $D_2$) levels at 847 keV and 2658 keV in $^{56}$Fe, respectively. As the spins and parities for the two final levels are equivalent, it is reasonable to  assume that the initial level density $\rho(E_i)$ and the population-depopulation factor $\sigma(E_i,J_i)g(E_i,J_i)$ of the initial levels that feed the final states in the diagonals are also the same.
Therefore, the number of counts in the diagonals for a given $E_i$ only needs to be corrected by the $E_\gamma^3$ factor.  
Following the sewing steps outlined above for the pairs of intensities for each $E_i$, the shape of the $\gamma$SF is obtained and compared to the results of the Oslo method in Fig.~\ref{fig:fg_gsf_56Fe}b. 

Due to the lack of neutron-resonance spacing data for $^{56}$Fe, as $^{55}$Fe is unstable, previous works have relied on systematics to obtain the slope of the NLD and $\gamma$SF~\cite{Larsen2013a,Larsen2017}. 
Comparing the previous results with those of the new Shape method, we can conclude that the two normalizations previously used are indeed reasonable. However, as there is only a $\sim 30$\% relative change in the estimated NLD at $S_n$ ($\rho(S_n)$ = 2.18(59) MeV$^{-1}$ and 2.87(68) MeV$^{-1}$) between the two normalizations, we are not in a position to confirm which normalization is correct. 
If there was a more pronounced discrepancy in slope between the different normalizations, the present method may enable a discrimination between the input spin-distribution models. 
Although the systematics used in $^{56}$Fe is appropriate there is no compelling reason to assume that systematic approaches can be extended to all nuclei. Hence, if no reliable systematics can be made, such as for nuclei far away from stability, the present method, which is based on a sound foundation, clearly provides a significant constraint on the slope of the NLD and $\gamma$SF. The low and high-energy discrepancies observed in Fig.~\ref{fig:fg_gsf_56Fe}b are further explored in Sec.~\ref{sec:level5}.

\subsection{Several Diagonals with different final $J^\pi$ combinations: $^{92}$Zr}

Data from the (p,p') reaction populating $^{92}$Zr \cite{Guttormsen2017} were used with the $\gamma$ rays detected in the NaI(Tl) CACTUS array \cite{Guttormsen1990} and the charged particles in SiRi. With $N=52$, $^{92}$Zr is close to the magic $N=50$ shell closure and is characterized by few low-lying levels. With the present experimental resolution it is possible to identify four diagonals. With the six combinations  $D_1D_2$, $D_1D_3$, $D_1D_4$, $D_2D_3$, $D_2D_4$, and $D_3D_4$ one can investigate the consistency between the various $\gamma$SFs from the Shape and Oslo-method results.

\begin{figure*}[t]
    \includegraphics[clip,width=1.9\columnwidth]{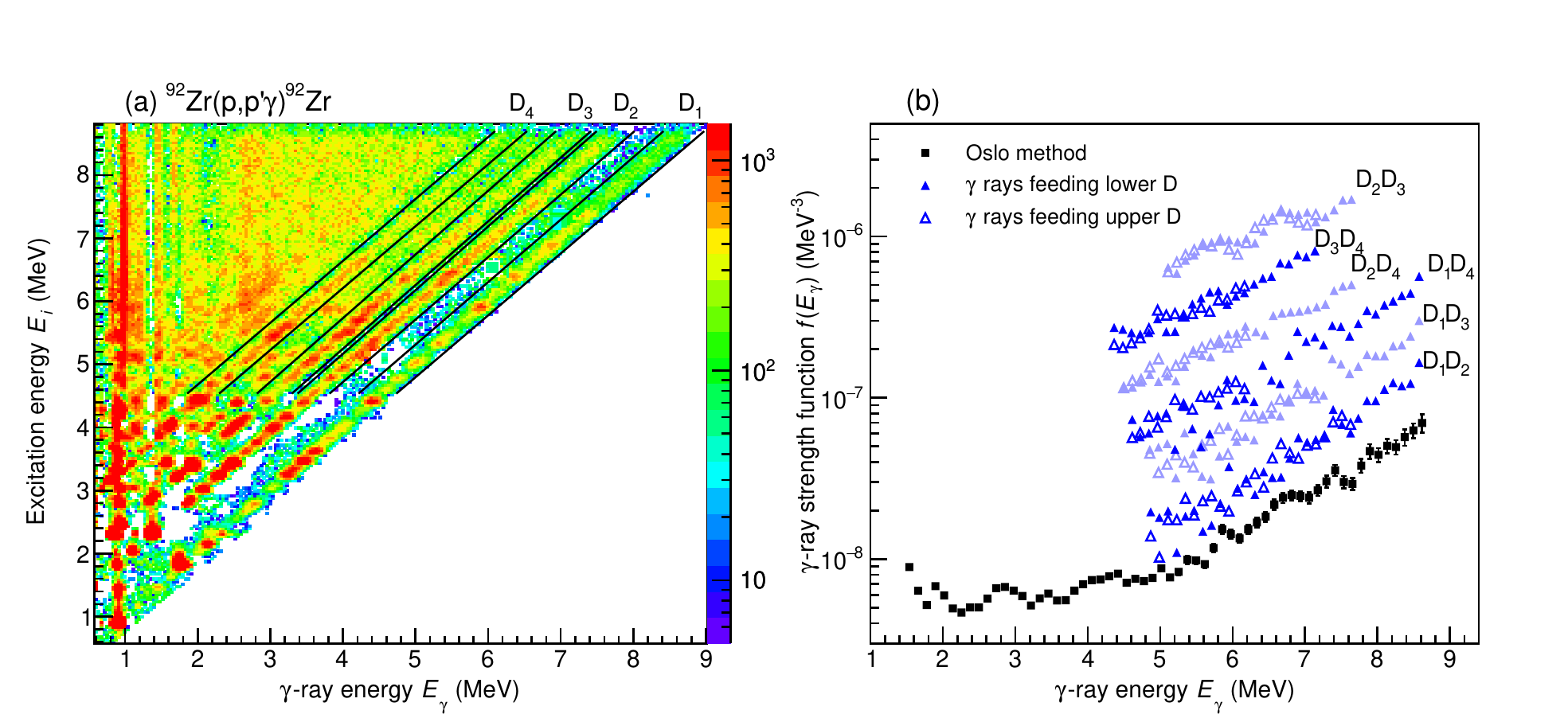}
    \caption{(Color online) (a) The first-generation matrix $P(E_{\gamma}, E_i)$  of $^{92}$Zr showing the four diagonals described in the text. (b) The resulting $\gamma$SFs from the Shape method (filled and open triangles in blue and light blue) compared to the Oslo-method results (black squares). The individual Shape method results are shifted in the plot in order to visualize the results from the various combinations of diagonals.}
    \label{fig:diablo_plot_92zr}
\end{figure*}

Figure \ref{fig:diablo_plot_92zr}a shows the primary matrix with the diagonals $D_j$ which include the following discrete states:
\begin{itemize}
\item[] $D_1$: 0$^+$(0 keV)
\item[] $D_2$: 2$^+$(934 keV)
\item[] $D_3$: 0$^+$(1383 keV) and 4$^+$(1495 keV)
\item[] $D_4$: 3$^-$(2340 keV), 4$^+$(2398 keV), and 5$^-$(2486 keV).
\end{itemize}
The lower part of the matrix shows that many non-statistical $\gamma$-ray transitions connect discrete levels and it is important to point out that these should not be taken into account when extracting the average $\gamma$SF for $^{92}$Zr. Thus, the results for the Oslo method in Fig.~\ref{fig:diablo_plot_92zr}b was extracted for $E_i > 4.5$~MeV. 

The same caution should be taken when applying the Shape method with the requirement that the final levels are well-defined states such as the ground state or first-excited states. Moreover, to maintain the statistical properties there should be enough initial states within the energy bin at $E_i$ that feed the levels contained by the diagonals. 
For $^{92}$Zr we obtain erratic fluctuations for $E_{\gamma} < 5$~MeV and this data is not shown. 

It is gratifying that the six extracted $\gamma$SFs from the Shape method are all in rather good agreement with the functional form between each other and the one obtained with the Oslo method. Since the combination of diagonals represent a variety of final $J^\pi$ values, yet they provide consistent functional forms, the spin distribution $g(E,J)$ applied in Eq.~(\ref{eq:spindist}) with spin cutoff parameters of Table~\ref{tab:parameters} is supported.

\subsection{Diagonals including Ground and Two-Quasiparticle Bands: $^{164}$Dy}

\begin{figure}[h]
    \includegraphics[clip,width=0.9\columnwidth]{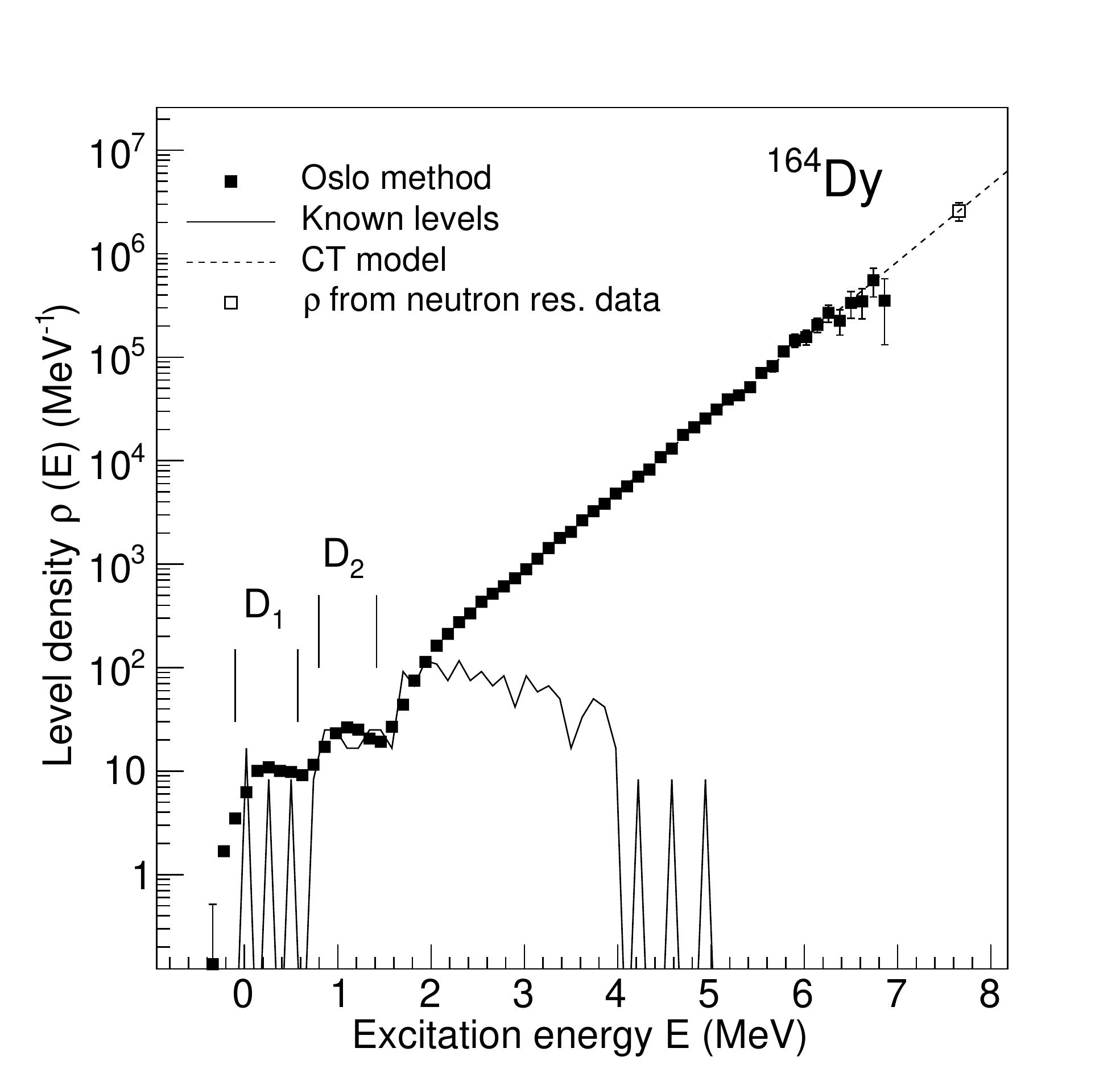}
    \caption{ Level densities of $^{164}$Dy \cite{Nyhus2012}. The solid line represents the NLD of known levels. The filled square symbols show the results of the Oslo method. The data points are connected to the NLD at $S_n$ (open square) through extrapolation with the constant temperature (CT) model.}
    \label{fig:counting164dy}
\end{figure}

\begin{figure*}[t]
    \includegraphics[clip,width=1.9\columnwidth]{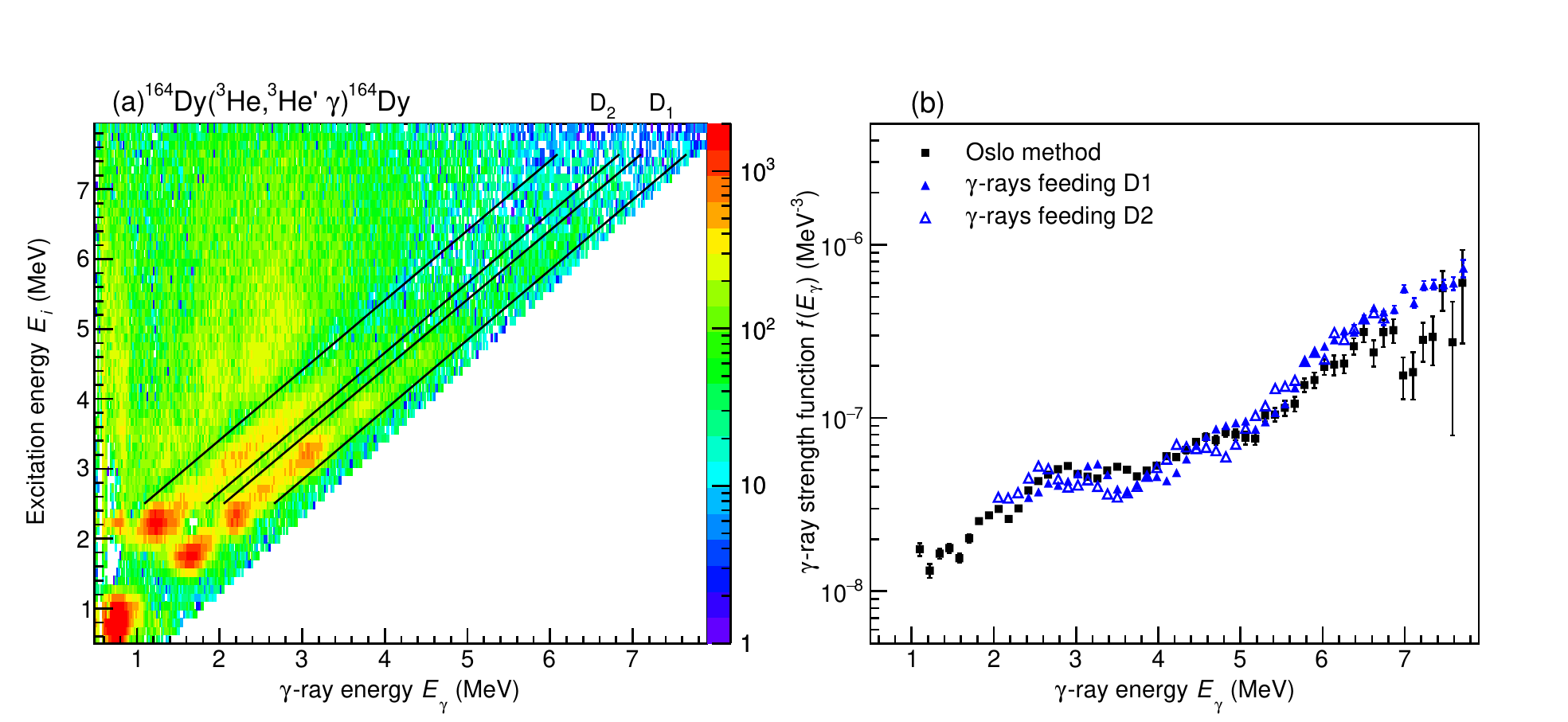}
    \caption{(Color online) (a) The first-generation matrix $P(E_{\gamma}, E_i)$  of $^{164}$Dy showing the two diagonals described in the text. (b) The resulting $\gamma$SF from the Shape method (filled and open blue triangles) compared to the Oslo-method results (black squares).}
    \label{fig:diablo_plot_164dy}
\end{figure*}

For rare earth nuclei the level density becomes high enough that it is difficult to identify final levels in the $P(E_{\gamma}, E)$ matrix within the experimental resolutions. However, the known levels of
$^{164}$Dy group into the ground band between $0-0.5$ MeV and two-quasiparticle band structures around 1.1 MeV. Figure~\ref{fig:counting164dy}
illustrates the level density obtained with the Oslo method which displays these two relatively well-defined structures. This makes  $^{164}$Dy a feasible case for applying the Shape method to the $^{164}$Dy$(^3$He,$^3$He') experimental data, measured with the CACTUS and SiRi arrays, from Refs. \cite{Nyhus2010, Nyhus2012, Renstrom2018}. Furthermore, there are two interesting features in the previous findings of the $\gamma$SF: ({\em i}) a scissors resonance at $E_{\gamma}= 2.83(8)$~MeV is built on the tail of the giant dipole resonance and ({\em ii}) it has been speculated if an enhancement exists around $E_{\gamma}= 6-7$~MeV due to the $E1$-pygmy resonance \cite{Renstrom2018}.
From the matrix in Fig.~\ref{fig:diablo_plot_164dy}a we immediately recognise the diagonals corresponding to the ground and two-quasiparticle bands by inspecting the distribution of known levels. Here, diagonal $D_1$ includes the 0$^+$, 2$^+$, 4$^+$ and 6$^+$ levels of the ground state band in the excitation region of $0-0.5$~MeV. Diagonal $D_2$ includes 14 levels in the excitation region of $0.76-1.39$~MeV, all with known $J^\pi$ \cite{NNDC}. Figure~\ref{fig:diablo_plot_164dy}b shows the $\gamma$SF extracted with the Oslo method \cite{Renstrom2018} together with the Shape method results.

It is interesting to note that the scissors resonance is directly visible from Fig.~\ref{fig:diablo_plot_164dy}a as a yellow-shaded region for $E_i> 4$~MeV and  $E_{\gamma}\sim 2-3$~MeV. This enhanced intensity is the main contributor to the scissors resonance strength obtained with the Oslo method. It is therefore rather exciting that the same information is also contained in the two diagonals used in the Shape method, resulting in a similar enhancement for $E_{\gamma}\sim 2-3$~MeV. 

Furthermore, the Shape method provides data up to $S_n$ with an apparent deviation in slope at $E_{\gamma}\sim 5.5$~MeV which may signal the presence of a resonance located in the $E_{\gamma}\sim 6-7$~MeV region. The previous results using the Oslo method were hampered by low statistics at the highest energies, as indicated by the large uncertainties for $E_\gamma > 6.6$ MeV, and therefore did not allow for a strong statement regarding the existence of an enhancement \cite{Renstrom2018}.

\subsection{Diagonals with many final levels of different $J^\pi$: $^{240}$Pu}

The $^{240}$Pu isotope was populated in the (d, p) reaction with a beam energy of 12 MeV and the $\gamma$ rays detected with the CACTUS and charged particles with the SiRi arrays. The excitation energy range analyzed here was restricted to $E_i < 4.5$~MeV due to the onset of fission; a limit much lower than the neutron separation energy of $S_n=6.534$~MeV. Further details of the experimental set-up and considerations are given by Ref. \cite{zeiser2019} and all results presented here are based on a reanalysis of the data.

The low-spin transfer of this sub-Coulomb barrier reaction is responsible for the population of only a fraction of the total intrinsic levels.
An iterative procedure was developed \cite{zeiser2019} that aims to correct for the bias introduced in the Oslo method. The populated $J^\pi$ distributions were estimated by the Green’s function transfer formalism and applied in $\gamma$-decay simulations to obtain consistent results \cite{Potel2015, Potel2017, zeiser2019}. 
In the following, we explore the possibility to apply the Shape method, even though the calculated $J^\pi$  distribution may not fulfill the assumptions on $\sigma(E_i, J_i)$ specified in Sec.\,\ref{sec:level3}. 
If the Shape method can be used to reliably extract the slope of the $\gamma$SF, it would be significantly easier to apply it than the iterative procedure proposed in Ref.\,\cite{zeiser2019}.

\begin{figure}[h]
    \includegraphics[clip,width=0.9\columnwidth]{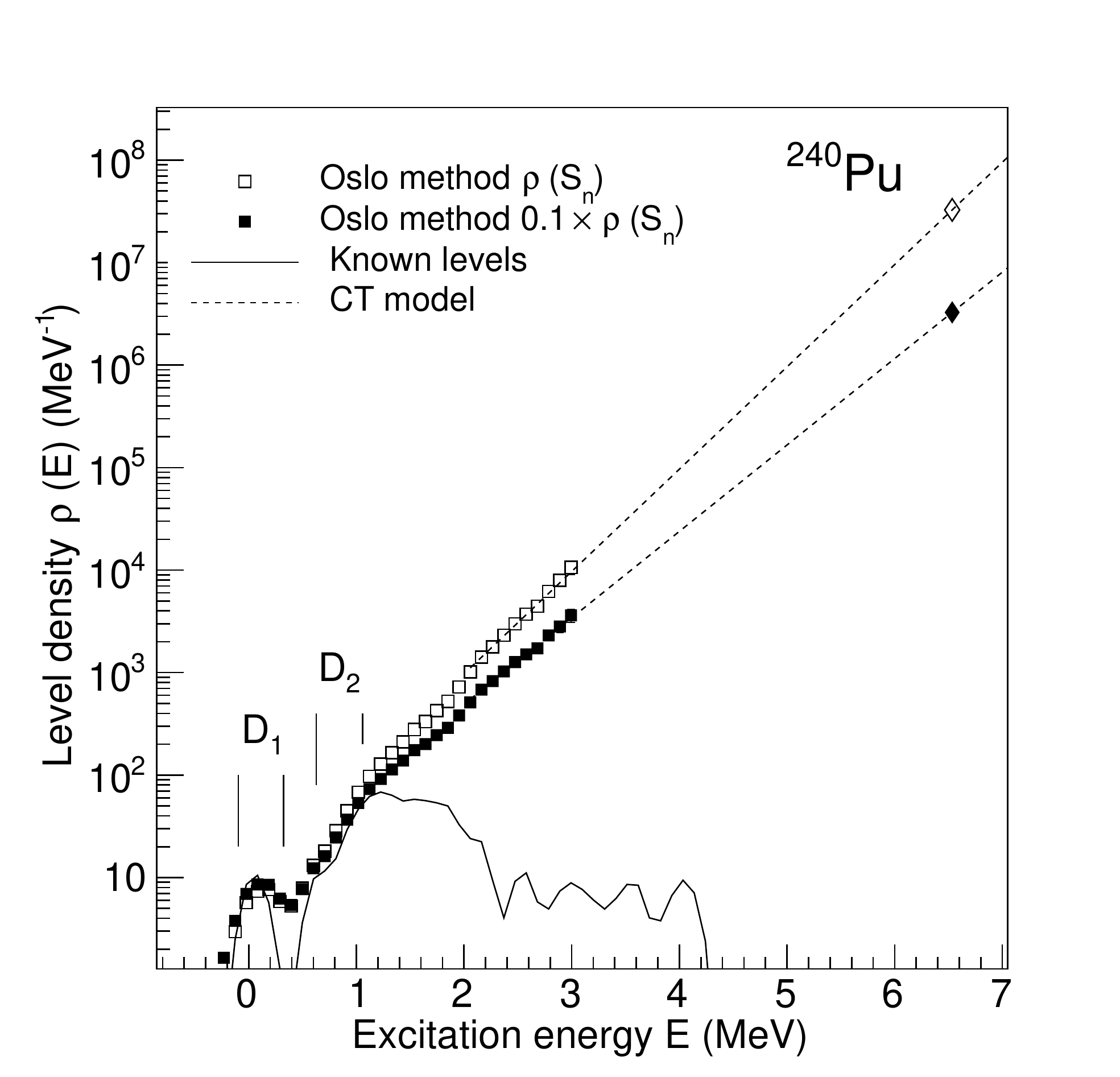}
    \caption{Level densities of $^{240}$Pu. The solid line shows the NLD of known levels. The open squares represent the results of the Oslo method when $\rho$ is normalized to the total level density at $S_n$, while the filled squares show the results when the reduced population of high-spin levels is taken into account. The reduction factor $0.1$ is obtained from a comparison of the $\gamma$SFs from the Oslo and the Shape method. The data points are extrapolated to the corresponding NLDs at $S_n$ with a constant temperature (CT) model (dashed lines). Note that the error bars are less than the size of the data points.}
    \label{fig:counting240pu}
\end{figure}

\begin{figure*}[t]
    \includegraphics[clip,width=1.9\columnwidth]{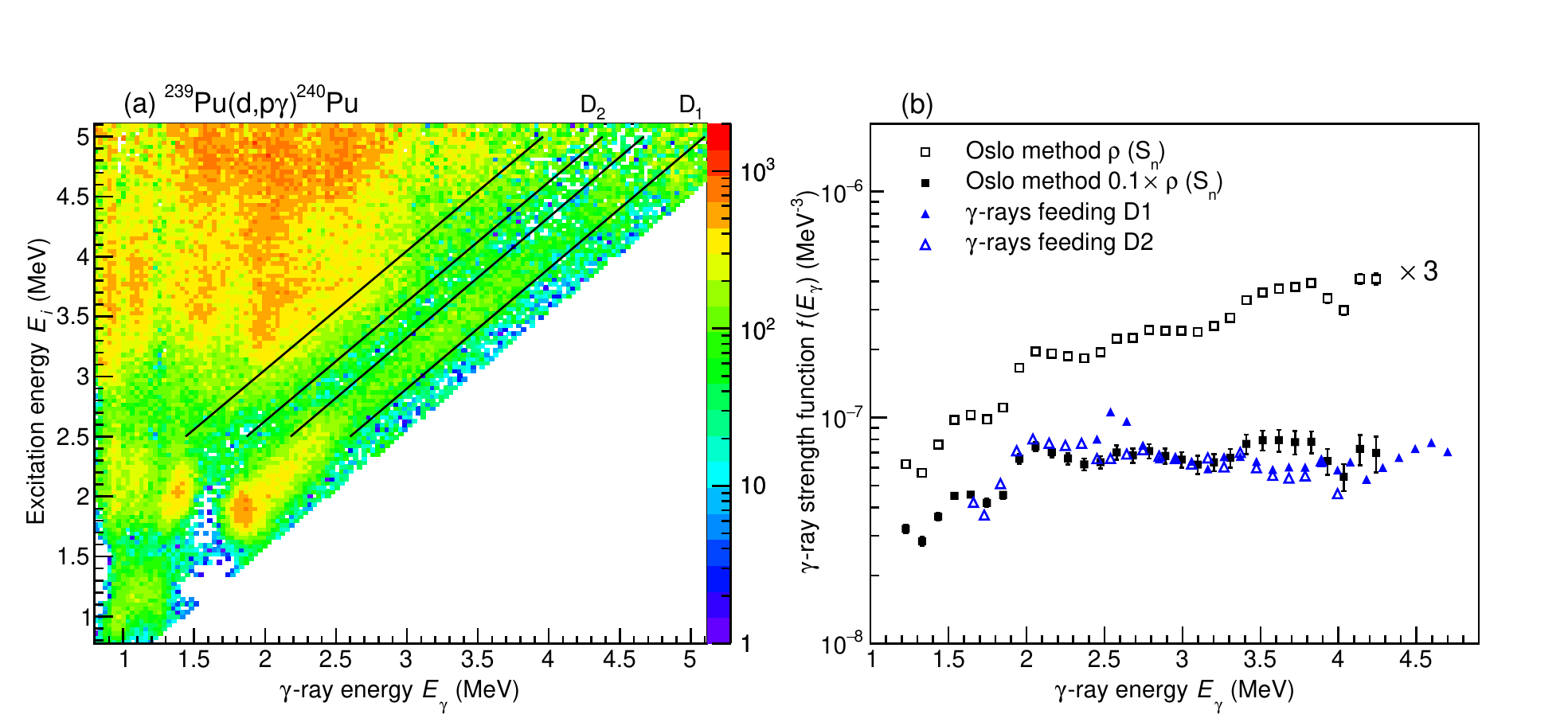}
    \caption{(Color online) (a) The first-generation matrix $P(E_{\gamma}, E_i)$  of $^{240}$Pu showing the two diagonals described in the text. (b) The $\gamma$SF obtained with the Shape method (filled and open blue triangles) compared to the Oslo-method results with a reduced spin range (solid black squares). The results with a full spin distribution are shown as open black squares and multiplied by a factor of 3 to facilitate readability of the figure.} \label{fig:diablo_plot_240pu}
\end{figure*}

A reduced spin population may be a challenge for the Oslo method since it is not clear what effect a varying $J^\pi$ population $\sigma(E_i, J_i)$ has on the first-generation method \cite{Larsen2011, zeiser2019, Zeiser2018a}. Nonetheless, we will now assume that $\sigma(E_i, J_i)$ does not significantly impact the overall results of the first-generation matrix $P(E_{\gamma}, E_i)$. To account for the fact that high-spin levels are rarely populated in the sub-Coulomb barrier reaction, the level density $\rho(S_n)$ used in the decomposition of $P(E_{\gamma}, E_i)$, see Eq.\,(\ref{102}) and Eq.\,(\ref{105}), has to be reduced by a factor $r$. This factor is directly linked to the slope of the $\gamma$SF through the normalization Eqs.\,(\ref{105}) and (\ref{106}), such that it can be determined by a comparison of the $\gamma$SF from the Oslo and the Shape methods. 

The key for an investigation with the Shape method is to identify two diagonals in the $P(E_{\gamma}, E_i)$ matrix which include a known number of final levels with proper spin assignments. 
The $^{240}$Pu isotope is one of the best studied nuclei in this mass region with a complete level scheme up to $\approx$~1~MeV. Figure~\ref{fig:counting240pu} shows the known levels of spin $J < 5$, which can be used to define the diagonals. The diagonal $D_1$ includes the first  $0^+$, $2^+$ and $4^+$ levels with an average final excitation energy of $E_f=62$~keV. The second diagonal $D_2$ has nine levels between $0.6 - 1.0$~MeV with an average energy of $E_f=849$~keV and an average spin of 2.3 $\hbar$. Figure~\ref{fig:diablo_plot_240pu}a shows the two diagonals chosen and the resulting $\gamma$SF pairs are presented in Fig.~\ref{fig:diablo_plot_240pu}b. 
The slope of the $\gamma$SF obtained with the Shape method is in agreement to the slope obtained with the Oslo method when a reduction factor of $\approx 0.1$ is applied to $\rho(S_n)$. 
We also display the $\gamma$SF if one assumes that all spins are populated in the reaction, which displays a significantly steeper slope. The corresponding NLDs used to extract the $\gamma$SFs are shown in Fig.~\ref{fig:counting240pu}.

It is difficult to make rigorous conclusions on the reduction factor $r$ since diagonal $D_2$ may have missing levels. In addition, there are uncertainties at the upper limit of 1 MeV to determine which levels are included within the experimental detection resolution. Thus, the case of $^{240}$Pu is meant to highlight the possibilities that may exist if reliable diagonals can be defined with high experimental resolution. 

It can be seen that the resulting $\gamma$SF is relatively flat between $E_\gamma \sim 2 - 4.5$ MeV. Further investigations are needed to probe whether this feature is due to the assumptions on $\sigma(E_i, J_i)$, or whether there is a strong enough contribution of e.g. the scissors resonance between 2 and 4 MeV that leads to an almost constant tail of the $\gamma$SF within the narrow $E_\gamma$ range considered.

\section{Discussion}
\label{sec:level5}

The shapes of the $\gamma$SFs extracted with the Oslo method are well reproduced with the  Shape method, in particular for excitation energies for which the total NLD of initial states is high. 
With reduced excitation energies discrete structures may become dominant and the concepts of $\gamma$SF and NLD are no longer applicable. 
This situation is apparent when inspecting the $\gamma$SF of $^{56}$Fe in Fig.~\ref{fig:fg_gsf_56Fe}b where the $\gamma$SF below $E_\gamma\sim 5.5$ MeV ($E_i \sim 6.5$ MeV) exhibits significant fluctuations. 
The NLD at $E_i = 6 $ MeV has been measured to be $\rho \sim 100~ \mathrm{MeV}^{-1}$ \cite{Algin2008}.
For $^{92}$Zr the Shape method has been applied from $E_i = $ 4.5 MeV where $\rho \sim 180~ \mathrm{MeV}^{-1}$ \cite{Guttormsen2017}. For the heavier nucleus $^{164}$Dy the level density reaches $\rho \sim 800~ \mathrm{MeV}^{-1}$ at $E_i = 3$ MeV and for $^{240}$Pu $\rho \sim 1000~ \mathrm{MeV}^{-1}$ at $E_i=2.5$ MeV as evident from Figs.~\ref{fig:counting164dy} and ~\ref{fig:counting240pu}, respectively. The relatively high NLDs found in $^{164}$Dy and $^{240}$Pu allow for the Shape method to be applied to low enough $E_i$ values to cover the range of the scissors resonance. It is important to emphasize that careful considerations have to be given to identify appropriate $E_i$ regions for the Shape method to be applicable. Discrete states and/or structures may become dominant features which lie outside the statistical regime. This is particularly the case for light $A$ nuclei or those which are located near closed shells. From our investigation, a minimum of $\rho \sim 100
~\mathrm{MeV}^{-1}$ appears to be appropriate, or more specifically, one should have more than $\approx 10$ transitions connecting the initial and final excitation energy bins. It is nonetheless recommended that each nucleus is being investigated carefully to determine the lowest reliable $E_i$ and hence lowest $\gamma$-ray energy to be used. 

At higher $E_i$, the data points from the Shape method follow the functional form of the $\gamma$SFs from the Oslo method rather well. At the highest $E_i$, the Oslo method may  underestimate the $\gamma$SF due to reduced statistics whereas the Shape method remains robust in this regime.\footnote{For $^{56}$Fe the low statistics is due to the very few levels up to $E_i \approx$ 3 MeV, which leads to a depletion of counts at high energies of the $\gamma$SF. For  $^{164}$Dy the matrix has low statistics at high energies as indicated by the large uncertainties.} As demonstrated for the four nuclei under consideration, it is in the region of higher $\gamma$-ray energies where the slope of the $\gamma$SF can be reliably obtained with the Shape method and provides the necessary constraints if alternative normalization procedures are not possible due to the absence of neutron resonance data.

Nuclei such as $^{56}$Fe, for which two low-lying discrete states of the same $J^{\pi}$ can be separated experimentally, represent the most fundamental application of the Shape method and can be treated with the fewest assumptions and without any model input. In such cases, the NLD and cross section dependencies of primary transitions feeding the states are eliminated. 

The Shape method remains applicable even when the discrete levels differ in $J^{\pi}$ or if the states cannot be resolved experimentally. This is clearly demonstrated for $^{92}$Zr where six different combinations of final levels all yield strikingly similar functional forms of the $\gamma$SF. This illustrates the robustness of the applied spin distributions and the assumption that the population cross-section is proportional to the spin distribution over the $E_i$ ranges considered for the extraction of $\gamma$SF below the particle thresholds. 

The results from $^{164}$Dy further reveal that the inclusion of many final levels of widely varying $J^\pi$ values or even distinctive nuclear structures still leads to an energy dependence which is in agreement with that of the $\gamma$SFs from the Oslo method. The $^{164}$Dy Oslo method results show the presence of the scissors resonance. The same information is retained in both diagonals and the resonance is reproduced by the Shape method. This may imply that this resonance is a collective mode obeying the Brink-Axel hypothesis. A suspected pygmy resonance at $E_\gamma \sim 6 - 7 $ MeV is apparent through the changing slope in $^{164}$Dy, while previous results were inconclusive \cite{Renstrom2018}, highlighting the complementary nature of the Shape method.

$^{240}$Pu represents an extreme case due to the reaction proceeding below the Coulomb barrier yielding a very limited spin-distribution. This requires $\rho(S_n)$ to be modified through the Oslo method, which propagates to the normalization of the $\gamma$SF, in order to reproduce the Shape method results. It is important to note, once the appropriate corrections are performed that both methods yield a similar energy dependence of the $\gamma$SFs despite the selectivity of the reaction. The reduced strength at $E_\gamma \sim$ 3.7 MeV from the Shape method may be indicative of a feature which  depends on the population/reaction mechanism. 

It is interesting to note that the results from the Shape method clearly yield very similar $\gamma$SFs, regardless if the $\gamma$SFs are built on different nuclear structures or $J^\pi$ states of a given nucleus. This confirms the validity of the generalized Brink-Axel hypothesis, supporting previous results \cite{Guttormsen2016}.        
Another appealing aspect of the Shape method is the fact that it can be applied to the same set of experimental data as that used to extract the NLD and $\gamma$SF with the Oslo method. This is highly beneficial when the Shape method is used to specifically determine the slope for the NLD and $\gamma$SF from the Oslo method since it avoids unnecessary additional systematic uncertainties which would arise when performing different experiments.

\section{Summary}
\label{sec:level6}
It has long been a challenging endeavour to estimate the slope of the $\gamma$SF in the absence of neutron resonance data which is compounded by the fact that no standardized approach exists which is applicable to all nuclei. The Shape method provides a solution to the $\gamma$SF normalization conundrum when $D_0$ values are not available. It provides a standardized approach to determine the slope of the $\gamma$SF and NLD (if extracted simultaneously through the Oslo method), which is not only universally applicable but will also provide consistency for analyses and results. 

The Shape method makes use of concepts from the Average Resonance Proton Capture, Ratio, and $\chi^2$ methods and is based on the unambiguous experimental identification of the origin and destination of primary $\gamma$-ray transitions. Through their intensities, pairs of primary transitions retain the information on the functional form of the $\gamma$SF. 

The Shape method has been applied to four nuclei which are representative of the various situations encountered: i) low-mass $^{56}$Fe, ii) $^{92}$Zr located in the vicinity of shell closures, iii) $^{164}$Dy with scissors and pygmy resonances, and iv) high-$Z$ nucleus $^{240}$Pu where the reaction proceeds below the Coulomb barrier. These four nuclei further represent a variety of  $J^\pi$ combinations for low-lying states which are fed by the primary transitions. 

In $^{56}$Fe, the primary transitions feed two well-separated and experimentally-resolved states of the same $J^\pi$, while in $^{92}$Zr some of the low-lying states cannot be resolved and are of different $J^\pi$. For $^{164}$Dy the low-lying states can only be identified through clusters of specific nuclear structures in the form of the ground and two-quasiparticle bands. The $^{240}$Pu case has an even larger number of final states which cannot be resolved experimentally. Regardless of the intricacies and details of the individual nuclei considered, the Shape method extracts functional forms of $\gamma$SFs which are consistent with those from the Oslo method. This highlights the robustness of the method and, where applicable, the appropriateness of the assumptions made regarding the spin distributions. While the Shape method provides a universal prescription to determine the slope of the $\gamma$SF (and for the NLD in the case of the Oslo method) in the absence of experimentally measured neutron resonance spacing it does not provide the absolute values of the $\gamma$SFs when neutron resonance widths are not available. Further work is highly desirable to explore alternate approaches to determine the absolute values of $\gamma$SFs.

Complementary to this work, we have also applied the Shape method to  $^{76}$Ge and $^{88}$Kr for the extraction of model-independent nuclear level densities away from stability \cite{Mucher2021}.

\begin{acknowledgements}
This work is based on the research supported in part by the National Research Foundation of South Africa (Grant Number: 118846), by the Research Council of Norway (Grant Number: 263030), and the National Science Foundation (Grant Number: PHY 1913554).
A.~C.~L. acknowledges funding of this research by the European Research Council through ERC-STG-2014 under grant agreement no. 637686, support from the “ChETEC” COST Action (CA16117), COST (European Cooperation in Science and Technology), and from JINA-CEE through the National Science Foundation under Grant No. PHY-1430152 (JINA Center for the Evolution of the Elements). 
\end{acknowledgements}

\end{document}